\RequirePackage{fix-cm}
\documentclass[smallextended]{svjour3}       
\smartqed  

\usepackage{graphicx}
\usepackage{amsmath,amssymb,amsthm}
\usepackage{comment}
\usepackage{caption}
\usepackage{booktabs}
\usepackage{hyperref}
\usepackage{float}
\usepackage{mathrsfs}  
\usepackage{tikz}
\usetikzlibrary{topaths,calc}
\usepackage[square,sort,comma,numbers]{natbib}
\begin{document}

\title{Hyperedge Prediction using Tensor Eigenvalue Decomposition
}


\author{Deepak Maurya        \and
        Balaraman Ravindran 
}


\institute{Computer Science and Engineering Dept., 
Robert Bosch Centre for Data Science and AI \at
              Indian Institute of Technology Madras, India \\ 
              \email{maurya@cse.iitm.ac.in, ravi@cse.iitm.ac.in}           
}

\date{Received: date / Accepted: date}

\maketitle

\begin{abstract}
Link prediction in graphs is studied by modeling the dyadic interactions among two nodes. The relationships can be more complex than simple dyadic interactions and could require the user to model super-dyadic associations among nodes. Such interactions can be modeled using a hypergraph, which is a generalization of a graph where a hyperedge can connect more than two nodes. 

In this work, we consider the problem of hyperedge prediction in a $k-$uniform hypergraph. We utilize the tensor-based representation of hypergraphs and propose a novel interpretation of the tensor eigenvectors. This is further used to propose a hyperedge prediction algorithm. The proposed algorithm utilizes the \textit{Fiedler} eigenvector computed using tensor eigenvalue decomposition of hypergraph Laplacian. The \textit{Fiedler} eigenvector is used to evaluate the construction cost of new hyperedges, which is further utilized to determine the most probable hyperedges to be constructed. The functioning and efficacy of the proposed method are illustrated using some example hypergraphs and a few real datasets. The code for the proposed method is available on \href{https://github.com/d-maurya/hypred_ tensorEVD}{this link}. 
\end{abstract}

\section{Introduction}
Link prediction is the study of predicting the existence of an edge between two nodes in a graph. This problem has found its application in various domains such as bioinformatics \cite{lu2011link,oyetunde2016boostgapfill}, social networks \cite{symeonidis}, and recommender systems \cite{koren2009matrix}. Most of existing heuristic approaches like common neighbours \cite{barabasi1999emergence}, Jaccard index \cite{liben2007link}, Adamic-Adar \cite{zhou2009predicting}, PageRank \cite{brin1998anatomy} are limited to modelling pairwise interactions only.

Relationships among nodes can be more complex than simple pairwise associations. Hypergraphs relax this assumption of pairwise interaction and provide the freedom to model the interaction among $k$ nodes. Such networks commonly occur in social networks \cite{li2013link, chodrow2021hypergraph}, metabolic networks \cite{zhang2018beyond}, recommender systems \cite{tarakci2014using, liu2020strongly} and multi-actor collaboration \cite{sharma2014predicting}. 

Most of the recent works on hyperedge prediction \cite{kumar2020hpra,Yadati2018LinkPI,zhang2018beyond} utilize the approach of clique expansion to reduce the hypergraph to a graph \cite{paper:agarwal_2006}, followed by applying standard graph algorithms \cite{zhou2007learning} or recently proposed approach modeling triadic simplicial closure \cite{benson2018simplicial}. We believe this step of hypergraph reduction to a graph restricts the user to model some weighted form of dyadic interaction rather than the intended super-dyadic interactions among nodes. This can be argued by the fact that the expression for scalar Laplacian objective function minimized for the estimation of Fiedler vector (eigenvector corresponding to minimum positive eigenvalue) contains only bi-linear terms  instead of higher order polynomials. This article emphasis on the use of higher order polynomials to capture the complex interactions among the nodes in a hypergraph. 

The availability of vast literature, theoretical guarantees, and scalable algorithms for graphs are the main reasons for adopting hypergraph reduction methods \cite{ghoshdastidar2017uniform}. There is significant loss of information about the hypergraph structure caused by this reduction step. For example, two entirely different hypergraphs can reduce to same graph after the reduction step. We have given example for such cases later. This observation clearly demonstrates that the unique information about two different hypergraphs is lost after the hypergraph reduction step. 

In this work, we approach the problem of hyperlink prediction without performing any reduction. For this purpose, we prefer to utilize the tensor-based representation of hypergraphs \cite{qi2017tensor} rather than the usual matrix based notation widely accepted in machine learning community \cite{zhou2004learning}. The tensor-based representation provides us the freedom to model the super-dyadic interactions among nodes. The proposed algorithm in this work is highly motivated from spectral graph theory with appropriate modifications. 

The widely accepted framework for link prediction using spectral graph theory framework comprises of computing the similarity metric between two nodes \cite{symeonidis}. The similarity measure is defined as a function of embeddings of two nodes, which can be derived from the graph Laplacian \cite{von2007tutorial}. We attempt to utilize the same approach for hypergraphs by computing the eigenvectors of hypergraph Laplacian tensor but we encountered several challenges. To name a few, the eigenvectors of a real symmetric tensor are not orthogonal, which is contrary to the case of real symmetric matrices. The number of eigenvectors for a symmetric tensor is not fixed, unlike the simple case of symmetric matrices. The tensor eigenvectors cannot be trivially interpreted. 

Despite various existing challenges, we pursue the tensor-based representation of hypergraphs due to the strong motivation developed from some of the recent intriguing results found in spectral hypergraph theory using tensor representation. For example, Hu \textit{et. al} \cite{hu2014eigenvectors} proved that the algebraic multiplicity of \textit{zero eigenvalue} of a symmetric tensor is equal to the sum of the number of even-bipartite connected components and number of connected components, minus the number of singletons in the corresponding hypergraph. This information couldn't be revealed from the clique reduction methods and its variants \cite{paper:agarwal_2006}.  

In this work, we present a novel approach to interpret the tensor eigenvectors. This helps us to define the construction cost for new potential hyperedges in a given hypergraph. The next step in the proposed algorithm is to prefer the prediction of hyperedges with minimum construction cost. In the perspective of spectral hypergraph theory, the key idea of the proposed algorithm can also be perceived as the inclusion of new hyperedges such that there is a minimal perturbation in the ``smoothness'' of the hypergraph. In spectral graph theory, the ``smoothness'' of the graph is characterized by the Fiedler eigenvalue \cite{chung1997spectral}. The same analogy is utilized in this work also. The code for the proposed method can be accessed from \href{https://github.com/d-maurya/hypred_tensorEVD}{this link}\footnote{https://github.com/d-maurya/hypred\_tensorEVD}. 

The rest of the paper is organized as follows. We introduce the matrix and tensor-based notation of hypergraphs in Section \ref{sec:prelims}. We also discuss the merits of tensor-based notation in this section. The proposed algorithm for hyperedge prediction is described in Section \ref{sec:prop_algo}. The functioning and fruitful merits of the proposed algorithm is demonstrated in Section \ref{sec:Exp} using small synthetic and real hypergraphs. Concluding remarks and future directions of this work are discussed in Section \ref{sec:concl}. 

{\bf Notations:} A scalar is denoted by lowercase alphabet $x$, a vector by bold face $\mathbf{x}$, a matrix by bold face uppercase alphabet $\mathbf{X}$ and a tensor by italics uppercase alphabet $\mathcal{X}$. The subscript $a$ over a vector such as $\mathbf{x}_a$ indicates the dimension, and for a tensor $\mathcal{X}_a$, it denotes the mode of a tensor, which is defined later.

\section{Preliminaries}
\label{sec:prelims}
In this section, we discuss the matrix and tensor-based representation of hypergraphs briefly. 

A hypergraph $G$ is formally defined as a pair of $G = (V, E)$, where $V = \{v_1, v_2, \ldots, v_n \}$ is the set of entities called vertices or nodes and $E = \{e_1, e_2, \ldots, e_m\}$ is a set  of non-empty subsets of $V$ referred as hyperedges. In general, $E \in  \mathcal{P}(V) \backslash \{\phi \}$ for a {\it non-uniform} hypergraph, where $\mathcal{P}(V)$  is the power set of $V$. In this article, we focus on $k$-uniform hypergraphs, and hence $E$ is restricted to a subset of  all the ${n \choose k}$ combinations of elements from $V$, where $n = |V|$. The strength of interaction among nodes in the same hyperedge is quantified by the positive weight represented by $w_e = \{w_{e_1}, w_{e_2}, \ldots,  w_{e_m} \}$. 
The vertex-edge incidence matrix is denoted by $\mathbf{H}$ and has the dimension $|V| \times |E|$. The entry $h(i, j)$ is defined as:
\begin{align}
h(i, j) = 
\begin{cases}
1 & \text{if $ v_i \in e_j $}  \\
0 & \text{otherwise}
\end{cases}
\end{align}
The degree of node $v_i$ is defined by $d(v_i) = \sum_{e_j \in E} w_{e_j} h(i, j)$. $\mathbf{D}$ is a diagonal matrix with $D(i,i)=d(v_i)$. 
\subsection{Matrix Representation}
It should be noted that there is no loss of information about the hypergraph structure in the incidence matrix representation. This could also be perceived as the existence of unique mapping  (up to certain reordering) between given hypergraph $G(V, E)$ and its incidence matrix $\mathbf{H}$. 

Most of the machine learning algorithms for classification, partitioning of graphs, and link prediction operates on the adjacency matrix rather than incidence matrix. The adjacency matrix for an undirected graph is symmetric and its entries $\mathbf{A}(i,j)$ is defined as:
\begin{align}
\mathbf{A}(i, j) = 
\begin{cases}
w_{ij} & \text{if $ \{v_i, v_j\} \in E $}  \\
0 & \text{otherwise}
\end{cases}
\end{align}
where $w_{ij}$ indicates the weight of the edge. It can be inferred that the entry $(i,j)$ denotes the existence of an edge. 

Agarwal \textit{et al.} \cite{paper:agarwal_2006} define the adjacency matrix for the reduced hypergraphs as follows:  
\begin{align}
    \mathbf{A}_r = \mathbf{H W} \mathbf{H}^T - \mathbf{D} \label{eq:inci_a}
\end{align}

\begin{figure}[htbp]
    \centering
    \includegraphics[scale = 0.3]{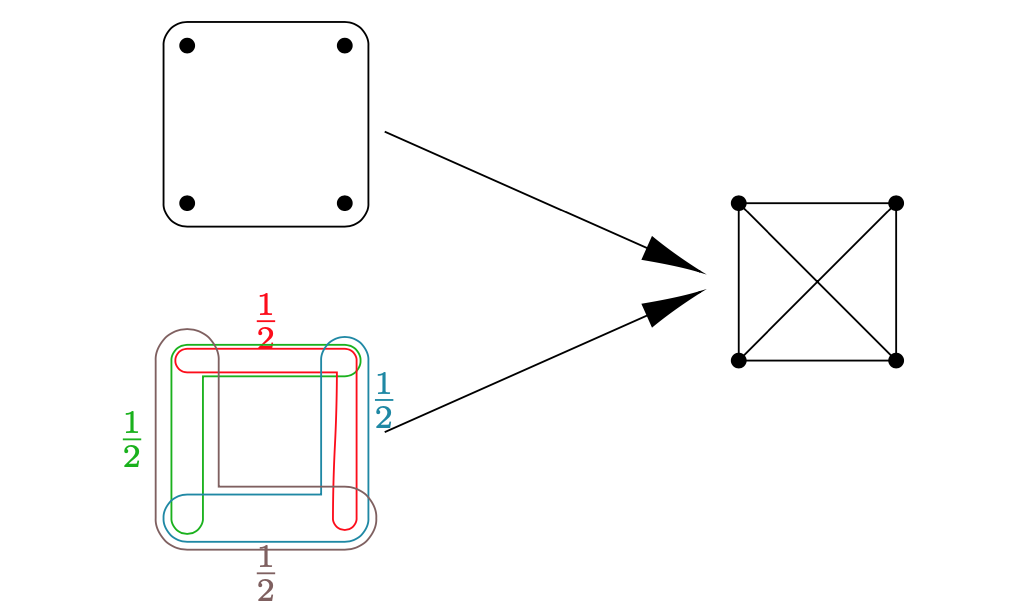}
    \captionsetup{justification=centering}
    \caption{Two Hypergraphs reducing to same Graph}
    \label{fig:2hy_sameGr}
\end{figure}

It should be noticed that most of the reduction methods are a non-unique mapping from hypergraph to adjacency matrix. It means that there could be distinct hypergraphs which reduce to the same graph. For example, the clique reduction approach reduces the four-uniform hypergraph and the three-uniform hypergraph hypergraph to the same graph as shown in Figure \ref{fig:2hy_sameGr}. 

This non-uniqueness property of hypergraph reduction method plays a very crucial role in the task of hyperedge prediction. The reduced hypergraph has lost the information about the original hypergraph structure. So there is no assurance of any analysis on reduced hypergraph to deliver correct results for the original hypergraph. To avoid the loss of information in the reduction step, we utilize the tensor-based representation of hypergraphs discussed in the next section. 

\subsection{Tensor-based Representation}
\noindent
The entry $\mathbf{A}(i,j)$ in adjacency matrix for a graph denotes the strength of interaction between the nodes $i$ and $j$. Similarly, for a hyperedge with $3$ nodes, a 3-dimensional tensor is required to represent the strength of interaction among $3$ nodes. This idea can be further generalized to represent $k$-uniform hypergraphs. 

Therefore, a natural representation of hypergraphs is a $k$-order $n$-dimensional tensor $\mathcal{A}$ \cite{qi2017tensor}, which consists of $n^{k}$ entries 
\begin{align}
	\mathcal{A} = \left( a_{i_1i_2\ldots i_k} \right), \qquad a_{i_1i_2\ldots i_k} \in \mathbb{R},  \qquad 1 \leq i_1, \ldots, i_k \leq n
\end{align}
The entries of above tensor is defined as:
\begin{align}
	a_{i_1 i_2\ldots i_k} = 
	\begin{cases}
   w_{e_j}	\frac{1}{(k-1) !} & \text{if $(i_1, i_2, \ldots, i_k) = \{e_j\} \quad e_j \in E$}  \\
	0 & \text{otherwise}
	\end{cases}
	\label{eq:adj_ten_def}
\end{align}
It should be noted that $\mathcal{A}$ is a ``\emph{super-symmetric}" tensor i.e, for any permutation of the indices, $\mathcal{A}$ contains the same value:
\begin{align}
	a_{i_1i_2\ldots i_k} = a_{\sigma \left(i_1i_2\ldots i_k\right)} \nonumber
\end{align}
where $\sigma \left(i_1, i_2, \ldots i_k\right)$ denotes any permutation of the elements in the set $\{i_1, i_2, \ldots ,i_k \} $. 
The order or mode of tensor refer to hyperedge cardinality, which is $k$ for $\mathcal{A}$.  
The degree of a vertex $v_i$ is given by 
\begin{align}
	d(v_i) = \sum_{i_k = 1}^{n} \ldots \sum_{i_3 = 1}^{n} \sum_{i_2 = 1}^{n} a_{ii_2i_3\ldots i_k}
\end{align}
The particular factor of $1/(k-1)!$ is chosen while defining the adjacency tensor in \eqref{eq:adj_ten_def} so that the node degree can be defined appropriately in the graph theoretic sense. One could also think as the generalization for the case of graphs (k=2). 

The degree of all the vertices can be represented by $k$-order $n$-dimensional diagonal tensor $\mathcal{D}$:
\begin{align}
d_{i_1 i_2\ldots i_k} = 
\begin{cases}
d(v_i) & \text{if $i_1 = i_2  \ldots = i_k = i $} \\
0 & \text{otherwise}
\end{cases}
\end{align}
The Laplacian tensor $\mathcal{L}$ is defined as follows : 
\begin{align}
	\mathcal{L} = \mathcal{D} - \mathcal{A} \label{eq:lap_t_den}
 \end{align}
The elements of Laplacian tensor $\mathcal{L}$ are described by 
\begin{align}
	l_{i_1 i_2\ldots i_k} = 
	\begin{cases}
	-w_{e_j}	\frac{1}{(k-1) !} & \text{if $(i_1, i_2, \ldots, i_k) \in \{e_j\},
	\quad j\in [m]$} \\
	d(v_i)  &\text{if $i_1 = i_2  \ldots = i_k = i $} \\ 
	0 & \text{otherwise}
	\end{cases}
	\label{eq:Lt_ent}
\end{align}

Normalized hypergraph Laplacian tensor \cite{banerjee2017spectra} denoted by $\mathscr{L}$, is defined: 
\begin{align}
\ell_{i_1 i_2\ldots i_k} = 
\begin{cases}
-w_{e_j}	\frac{1}{(k-1) !} \prod_{i_j=1}^{k} \frac{1}{\sqrt[k]{d_{i_j}}} & \text{if $(i_1, i_2, \ldots, i_k) \in \{e_j\}, \quad j \in [m]$} \\
1 & \text{if $i_1 = i_2  \ldots = i_k = i $} \\ 
0 & \text{otherwise}
\end{cases}
\label{eq:Lt_ent_norm}
\end{align}
One of the standard approaches in spectral graph theory is to perform spectral decomposition of the graph Laplacian for link prediction. Proceeding along similar lines, we evaluate the eigenvectors of Laplacian tensor \cite{qi2017tensor}: 
\begin{align}
	\mathcal{L} \mathbf{x}^{k-1} &= \lambda \mathbf{x} \nonumber \\
\mathbf{x}^T \mathbf{x} &= 1 \label{eq:ten_eig}
\end{align}
where $(\lambda, \mathbf{x}) \in (\mathbb{R}, \mathbb{R}^n \backslash \{0\}^n)$ is called the Z-eigenpair and $\mathcal{L} \mathbf{x}^{k-1} \in \mathbb{R}^n$, whose $i^{th}$ component  is defined as 
\begin{align}
	\left[ \mathcal{L} \mathbf{x}^{k-1} \right]_i = \sum_{i_k = 1}^{n} \ldots \sum_{i_3 = 1}^{n} \sum_{i_2 = 1}^{n} l_{ii_2i_3\ldots i_k} x_{i_2} x_{i_3} \ldots x_{i_k}  
\end{align}
The above equations arises from the following optimization problem: 
\begin{align}
\min_{\mathbf{x}} & \quad \mathcal{L} \mathbf{x}^{k} = \sum_{i_k = 1}^{n} \ldots \sum_{i_2 = 1}^{n} \sum_{i_1 = 1}^{n} l_{i_1i_2\ldots i_k} x_{i_1} x_{i_2} \ldots x_{i_k}  \label{eq:L_ful} \\
\text{such that}\quad  & \mathbf{x}^T \mathbf{x} = 1 \nonumber
\end{align}
The eigenvector with minimum positive  $\lambda $ satisfying \eqref{eq:ten_eig} is termed as Fiedler eigenvector and can be computed using 
\begin{align}
	\mathbf{v}_{\star} &= \underset{\mathbf{x}}{\mathrm{argmin}} \quad \mathcal{L}\mathbf{x}^{k} > 0 \label{eq:fiedler_eqn}\\ 
	\text{s. t}& \quad \mathcal{L}\mathbf{x}^{k} = \lambda \mathbf{x} \nonumber \\
	&\quad \mathbf{x}^T \mathbf{x} = 1  \nonumber
\end{align}
 The corresponding eigenvalue can be computed as $\lambda_{\star} = \mathcal{L}\mathbf{v}_{\star}^{k}$. 

\noindent
 {\bf Illustrative Example:}
\label{app:def_hyp_ex}
A simple example illustrating the procedure to construct the adjacency and Laplacian tensor for a $4-$uniform hypergraph has been presented. Consider a $4-$uniform hypergraph $G(V,E)$ as shown below:
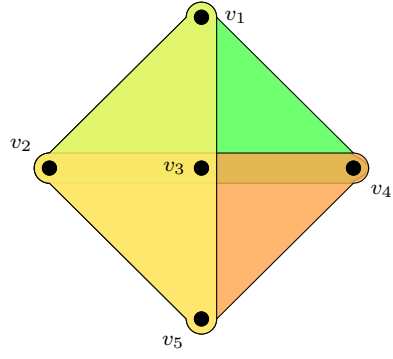
\begin{figure}[htbp]
	\centering
	\begin{tikzpicture}
    \node (v1) at (0,0) {};
    \node (v3) at (2,0) {};
    \node (v2) at (2,-2) {};
    \node (v9) at (2,2) {};
    \node (v4) at (4,0) {};
    
    \begin{scope}[fill opacity=0.8]
    \filldraw[fill=green!70] ($(v1)+(0,-0.2)$) 
        to[out=0,in=180] ($(v4) + (0,-0.2)$)
        to[out=180,in=270] ($(v4) + (0.2,0)$)
        to[out=90,in=180] ($(v4) + (0,0.2)$)
        to[out=135,in=-45] ($(v9) + (0.2,0)$)
        to[out=90,in=0] ($(v9) + (0,0.2)$)
        to[out=180,in=90] ($(v9) + (-0.2,0)$)
        to[out=225,in=45] ($(v1) + (0,0.2)$)
        to[out=180,in=90] ($(v1) + (-0.2,0)$)
        to[out=270,in=180] ($(v1) + (0,-0.2)$);
    \filldraw[fill=orange!70] ($(v1)+(0,0.2)$)
        to[out=0,in=180] ($(v4) + (0,0.2)$)
        to[out=0,in=90] ($(v4) + (0.2,0)$)
        to[out=270,in=0] ($(v4) + (0,-0.2)$)
        to[out=225,in=45] ($(v2) + (0.2,0)$)
        to[out=270,in=0] ($(v2) + (0,-0.2)$)
        to[out=180,in=270] ($(v2) + (-0.2,0)$)
        to[out=135,in=-45] ($(v1) + (0,-0.2)$)
        to[out=180,in=270] ($(v1) + (-0.2,0)$)
        to[out=90,in=180] ($(v1) + (0,0.2)$); 
    \filldraw[fill=yellow!70] ($(v2) + (0.2,0)$)
        to[out=270,in=0] ($(v2) + (0,-0.2)$)
        to[out=180,in=270] ($(v2) + (-0.2,0)$)
        to[out=135,in=-45] ($(v1) + (0,-0.2)$)
        to[out=180,in=270] ($(v1) + (-0.2,0)$)
        to[out=90,in=180] ($(v1) + (0,0.2)$)
        to[out=45,in=225] ($(v9) + (-0.2,0)$)
        to[out=90,in=180] ($(v9) + (0,0.2)$)
        to[out=0,in=90] ($(v9) + (0.2,0)$)
        to[out=270,in=90] ($(v2) + (0.2,0)$);
        
    \end{scope}

    \foreach \v in {1,2,3,4,9} {
        \fill (v\v) circle (0.1);
    }

    \fill (v1) circle (0.1) node [above left =0.12cm] {$v_2$};
    \fill (v2) circle (0.1) node [below left =0.12cm] {$v_5$};
    \fill (v3) circle (0.1) node [left = 0.1cm] {$v_3$};
    \fill (v4) circle (0.1) node [below right =0.12cm] {$v_4$};
    \fill (v9) circle (0.1) node [right =0.2cm] {$v_1$};
  
\end{tikzpicture}
	\caption{H4: 4-uniform }
	\label{fig:ex1_4uni}
\end{figure}
 
 \noindent where the set of vertices and hyperedges are defined by 
 \begin{align*}
 	V &= \{ v_1, v_2 ,v_3, v_4, v_5 \} \\
 	E &= \{ \{v_1,v_2,v_3,v_4\},  \{v_2,v_3,v_4, v_5\} , \{v_1,v_2,v_3, v_5\} \}
 \end{align*} 
 The adjacency tensor for the above hypergraph is denoted by $\mathcal{A}$ and has dimensions $5 \times 5 \times 5 \times 5$. It should be noted that the cardinality ($k$) of all $3$ hyperedges is 4 . The elements of $\mathcal{A}$ are denoted by $a_{i_1, i_2, i_3, i_4}$, where $1 \leq i_k \leq 5$. It should be noted that $\mathcal{A}$ contains $n^k = 5^4$ elements but only $m \times k! = 3 \times 4! = 72$ elements have non-zero entries. The elements corresponding to first hyperedge are described by 
 \begin{align*}
 	& a_{1234} = a_{1243} = a_{1324} = a_{1342} = a_{1423} = a_{1432} \\
 	= & a_{2134} = a_{2143} = a_{2314} = a_{2341} = a_{2413} = a_{2431} \\
 	= & a_{3214} = a_{3241} = a_{3124} = a_{3142} = a_{3421} = a_{3412} \\
 	= & a_{4231} = a_{4213} = a_{4321} = a_{4312} = a_{4123} = a_{4132} = c\\
 \end{align*}
 where $c = \frac{1}{(k-1)!} = \frac{1}{6}$. The vertex degrees can be stored in a tensor of dimension $5 \times 5 \times 5 \times 5$ with its diagonal elements being $d(v) = \begin{bmatrix}
 2 & 3 & 3 & 2 & 2
 \end{bmatrix}$.  The tensor Laplacian has dimension of $5 \times 5 \times 5 \times 5$ and its entries can be obtained using \eqref{eq:Lt_ent}, which are found to be 
\begin{align}
l_{i_1 i_2i_3 i_4} = 
\begin{cases}
-\frac{1}{6} & \text{if $(i_1, i_2, i_3, i_4) = \{e_j\}, \quad j= \{1,2,3\}$} \\
d(v_i) & \text{if $i_1 = i_2 = i_3 = i_4 = i $} \\ 
0 & \text{otherwise}
\end{cases}
\label{eq:L_ex1}
\end{align} 
This example shows the procedure to construct the adjacency and Laplacian tensor for any $k$-uniform hypergraph.

%

\section{Proposed Method for Hyperedge Prediction}
\label{sec:prop_algo}
In this section, we propose the hyperedge prediction algorithm using spectral decomposition for tensors. We have already discussed the tensor-based representation of hypergraphs in Section \ref{sec:prelims}. 

The proposed framework derives the cost of each potential hyperedge and prefers to choose the hyperedges with minimum cost. The cost of creating a hyperedge is calculated from the Fiedler eigenvector of the Laplacian tensor defined in \eqref{eq:fiedler_eqn}. Hence, we present the following theorem for the expression of hypergraph (tensor) Laplacian objective function used for the computation of tensor eigenvectors. 

\noindent
{\bf Theorem 1}: {\em The hypergraph Laplacian cost function for a k-uniform hypergraph can be expressed as }
\begin{align}
	\mathcal{L} \mathbf{x}^{k} &= \sum_{e_j \in E} l_{e_j}(\mathbf{x}) \nonumber \\ 
l_{e_j}(\mathbf{x})  &=	w_{e_j} \left( \sum_{i_k \in e_j } x^k_{i_k} - k \prod_{i_k \in e_j} x_{i_k} \right) \nonumber \\ 
	&= w_{e_j} k \left( \underset{ i_j \in e_j }{\text{A.M}\left(x^k_{i_k} \right)} -  \underset{ i_j \in e_j }{\text{G.M}\left(|x_{i_k}|^k \right)} (-1)^{n_s} \right) 
	\label{eq:am_gm_main}
\end{align}
where $n_s = |\{ i_j : x_{i_j} < 0 \}|$, A.M and G.M stand for the arithmetic and geometric means, respectively. We refer $l_{e_j}$ as the cost for hyperedge $e_j$ in rest of the paper.

\noindent
{\bf Proof: } Please refer Theorem 8 in Maurya \textit{et. al} \cite{maurya2020hypergraph}.

Using Theorem 1, the computation of $\mathcal{L} \mathbf{x}^{k}$ can be done in $O(|E|)$ steps which would have been $O(|V|^k)$ for any general tensor.

\noindent
{\bf Illustrative Example}: Through this example, we demonstrate the use of  Theorem 1 in the computation of tensor Laplacian. We also unveil the challenges involved in working with tensor eigenvectors — for example, the non-orthogonality of tensor eigenvectors. Consider the hypergraph shown in Figure \ref{fig:ex1_4uni}. The hypergraph Laplacian cost function for this hypergraph can be derived using \eqref{eq:am_gm_main}: 

\begin{align}
	\mathcal{L} \mathbf{x}^{k}  = \quad & x_1^4 + x_2^4 + x_3^4 + x_4^4 -4 x_1 x_2 x_3 x_4\nonumber \\
	     + & x_1^4 + x_2^4 + x_3^4 + x_5^4 -4 x_1 x_2 x_3 x_5 \nonumber \\
	     + &  x_2^4 + x_3^4 + x_4^4 + x_5^4 -4 x_2 x_3 x_4 x_5  
	     \label{eq:ex1_la_amgm} 
\end{align}
The $4^{th}$ order homogeneous polynomial is the objective function the optimization problem mentioned in \eqref{eq:L_ful}. This is required for the computation of the eigenvalues and eigenvectors of $\mathcal{L}$. We further discuss the properties of zero eigenvalues and zero eigenvectors of Laplacian tensor. 

\noindent 
{\bf Lemma 2:} {\em One of the Z-eigenpair of $\mathcal{L}$ is $(0, \mathbf{v})$, where $\mathbf{v} = \frac{1}{\sqrt{n}} \left( 1, 1, \ldots, 1\right) \in \mathbb{R}^n$. }

\noindent 
{\bf Proof:} 
Please refer Banerjee \textit{et. al} \cite{banerjee2017spectra}: Theorem 3.13 (iv). 
 \hfill $\Box$

It should be noted that $\sqrt{n}$ is just a scaling factor in $\mathbf{v}$ to ensure $\mathbf{v}^T \mathbf{v} = 1$. One could also consider unity vector as eigenvector with eigenvalue $0$.

\noindent 
{\bf Lemma 3:} {\em The cardinality of zero eigenvalues of the graph Laplacian indicates the number of connected components} \cite{chung1997spectral}.

The above property does not hold for hypergraphs. It means that a fully connected hypergraph can have multiple zero eigenvalues \cite{hu2014eigenvectors}. We also make the same observation in this example as explained below. 

It is observed that the tensor Laplacian for hypergraph H4 shown in Figure \ref{fig:ex1_4uni}  has $2$  zero eigenvalues and the distinct eigenvectors are as follows:
\begin{align}
	\mathbf{V} = \frac{1}{\sqrt{5}}
	\begin{bmatrix}
	1 & -1  \\
	1 & 1  \\
	1 & 1 \\
	1 & -1  \\
	1 & -1  \\
	\end{bmatrix}
\end{align}
This illustrates that the eigenvalues could be zero even if the eigenvector is not unity vector, contrary to graphs. This is surprising because there is only one connected component, but there are two zero eigenvalues. 

This observation can be explained by computing the cost of each hyperedge using \eqref{eq:am_gm_main} for the eigenvectors stated above. It is observed that the cost of all the three hyperedges is zero for both the eigenvectors. As a result of which, the eigenvalue is zero. Another distinguishable property is that the eigenvectors are \emph{not} orthogonal unlike the case of graphs (having real symmetric Laplacian matrix). 

We have just discussed the use of hyperedge cost from eigenvectors corresponding to zero eigenvalues. We extend the similar discussion on hyperedge score computed from the Fiedler eigenvalue and eigenvector. 

\noindent
{\bf Illustrative Example}: In this example, we demonstrate a novel interpretation of tensor eigenvectors. 
\begin{figure}[htbp]
	\centering
	\begin{tikzpicture}
    \node (v1) at (0,2) {};
    \node (v2) at (2,2) {};
    \node (v3) at (0,0) {};
    \node (v4) at (2,0) {};
    \node (v5) at (3,1) {};
    \node (v6) at (4,2) {};
    \node (v7) at (4,0) {};
    \node (v8) at (6,0) {};
    \node (v9) at (6,2) {};
    
    \begin{scope}[fill opacity=0.8]

    \draw[fill = yellow!70] ($(v1)+(-0.3,0)$) 
        to[out=270,in=90] ($(v3) + (-0.3,0)$) 
        to[out=270,in=180] ($(v3) + (0,-0.3)$)
        to[out=0,in=270] ($(v3) + (0.3,0)$)
        to[out=90,in=270] ($(v1) + (0.3,-0.3)$)
        to[out=0,in=180] ($(v2) + (0,-0.3)$)
        to[out=0,in=270] ($(v2) + (0.3,0)$)
        to[out=90,in=0] ($(v2) + (0,0.3)$) 
        to[out=180,in=0] ($(v1) + (0,0.3)$)
        to[out=180,in=90] ($(v1) + (-0.3,0)$) ; 

    \draw[fill = red!100] ($(v6)+(0,-0.2)$) 
        to[out=0,in=180] ($(v9) + (-0.2,-0.2)$) 
        to[out=270,in=90] ($(v8) + (-0.2,0)$)
        to[out=270,in=180] ($(v8) + (0,-0.2)$)
        to[out=0,in=270] ($(v8) + (0.2,0)$)
        to[out=90,in=270] ($(v9) + (0.2,0)$)
        to[out=90,in=0] ($(v9) + (0,0.2)$)
        to[out=180,in=0] ($(v6) + (0,0.2)$)
        to[out=180,in=90] ($(v6) + (-0.2,0)$)
        to[out=270,in=180] ($(v6) + (0,-0.2)$);
    
    \draw[fill = red!100] ($(v1)+(0,-0.2)$) 
        to[out=0,in=180] ($(v2) + (-0.2,-0.2)$) 
        to[out=270,in=90] ($(v4) + (-0.2,0)$)
        to[out=270,in=180] ($(v4) + (0,-0.2)$)
        to[out=0,in=270] ($(v4) + (0.2,0)$)
        to[out=90,in=270] ($(v2) + (0.2,0)$)
        to[out=90,in=0] ($(v2) + (0,0.2)$)
        to[out=180,in=0] ($(v1) + (0,0.2)$)
        to[out=180,in=90] ($(v1) + (-0.2,0)$)
        to[out=270,in=180] ($(v1) + (0,-0.2)$);
        
    \draw[fill = cyan!70] ($(v2) + (0.4,0)$) 
        to[out=270,in=90] ($(v4) + (0.4,0)$)
        to[out=270,in=0] ($(v4) + (0,-0.4)$)
        to[out=180,in=0] ($(v3) + (0,-0.4)$)
        to[out=180,in=270] ($(v3) + (-0.4,0)$)
        to[out=90,in=180] ($(v3) + (0,0.4)$)
        to[out=0,in=180] ($(v4) + (-0.4,0.4)$)
        to[out=90,in=90] ($(v2) + (-0.4,0)$)
        to[out=90,in=180] ($(v2) + (0,0.4)$)
        to[out=0,in=90] ($(v2) + (0.4,0)$);

    \draw[fill = green!100] ($(v4)+(-0.3,0)$) 
        to[out=45,in=225] ($(v6) + (-0.3,0)$) 
    to[out=45,in=90] ($(v6) + (0.3,0)$) 
    to[out=225,in=45] ($(v4) + (0.3,0)$) 
    to[out=250,in=225] ($(v4) + (-0.3,0)$);
    
   \draw[fill = yellow!70] ($(v9) + (0.4,0)$) 
        to[out=270,in=90] ($(v8) + (0.4,0)$)
        to[out=270,in=0] ($(v8) + (0,-0.4)$)
        to[out=180,in=0] ($(v7) + (0,-0.4)$)
        to[out=180,in=270] ($(v7) + (-0.4,0)$)
        to[out=90,in=180] ($(v7) + (0,0.4)$)
        to[out=0,in=180] ($(v8) + (-0.4,0.4)$)
        to[out=90,in=90] ($(v9) + (-0.4,0)$)
        to[out=90,in=180] ($(v9) + (0,0.4)$)
        to[out=0,in=90] ($(v9) + (0.4,0)$);
        
    \draw[fill = cyan!70] ($(v6) + (-0.25,0)$) 
        to[out=270,in=90] ($(v7) + (-0.25,0)$) 
        to[out=270,in=180] ($(v7) + (0,-0.25)$) 
        to[out=0,in=180] ($(v8) + (0,-0.25)$)
        to[out=0,in=270] ($(v8) + (0.25,0)$)
        to[out=90,in=0] ($(v8) + (0,0.25)$)
        to[out=180,in=0] ($(v7) + (0.25,0.25)$)
        to[out=90,in=270] ($(v6) + (0.25,0)$)
        to[out=90,in=0] ($(v6) + (0,0.25)$)
        to[out=180,in=90] ($(v6) + (-0.25,0)$);

    \end{scope}

    \foreach \v in {1,2,...,9} {
        \fill (v\v) circle (0.1);
    }

    \fill (v1) circle (0.1) node [above left =0.12cm] {$v_1$};
    \fill (v2) circle (0.1) node [above right = 0.2cm] {$v_2$};
    \fill (v3) circle (0.1) node [left = 0.3cm] {$v_3$};
    \fill (v4) circle (0.1) node [below= 0.35cm] {$v_4$};
    \fill (v5) circle (0.1) node [below right] {$v_5$};
    \fill (v6) circle (0.1) node [above = 0.3cm] {$v_6$};
    \fill (v7) circle (0.1) node [below= 0.33cm] {$v_7$};
    \fill (v8) circle (0.1) node [below = 0.33cm] {$v_8$};
    \fill (v9) circle (0.1) node [above = 0.3cm] {$v_9$};

\end{tikzpicture}

	\caption{H5: 3-uniform hypergraph}
	\label{fig:ex2_3uni}
\end{figure}
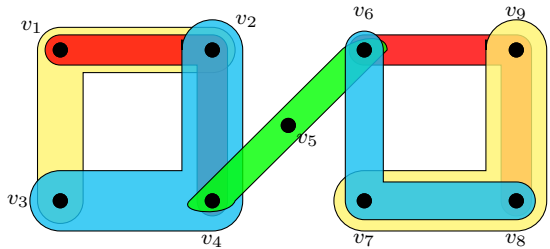

Consider the 3-uniform hypergraph shown in Figure \ref{fig:ex2_3uni}. The Laplacian tensor can be easily constructed using \eqref{eq:Lt_ent}. The next step is to compute the Fiedler eigenvalues and eigenvectors satisfying  \eqref{eq:ten_eig}. It is observed that there are four Fiedler vectors with the eigenvalue of $0.0569$ as reported below:
{\small
\begin{align}
    \mathbf{V} = 
    \begin{bmatrix}
        -0.05     &     0.06    &      0.47     &     0.47\\
         0.03     &     0.03    &      0.46     &     0.46\\
         0.06     &    -0.05    &      0.47     &     0.47\\
         0.23     &     0.23    &      0.42     &     0.42\\
         0.34     &     0.34    &      0.34     &     0.34\\
         0.42     &     0.42    &      0.23     &     0.23\\
         0.47     &     0.47    &     -0.05     &     0.06\\
         0.46     &     0.46    &      0.03     &     0.03\\
         0.47     &     0.47    &      0.06     &    -0.05 \\
    \end{bmatrix}
    \label{eq:ex1_eigvec}
\end{align}
}


We compute the cost of each hyperedge denoted by $l_{e_j}(\mathbf{x})$ in \eqref{eq:am_gm_main} using Fiedler vectors and tabulate in Table  \ref{tab:3uni_cosr}.  

\begin{table}[H]
	\caption{Hyperedge cost for Figure \ref{fig:ex2_3uni}}
	\label{tab:3uni_cosr}
\begin{center}
{\small \begin{tabular}{ |p{1.3cm}||p{1cm}|p{1cm}|p{1cm}| p{1cm}| }
	\hline
	Hyperedges    & $\mathbf{v}_1$ &$\mathbf{v}_2$&$\mathbf{v}_3$ & $\mathbf{v}_4$ \\
	\hline
	$\{ 1,2,3\}$   & $ 0.0004$ &$ 0.0004$&$ 0$ & $ 0$ \\
	$\{1,2,4\}$   & $ 0.0127$ &   $ 0.0111  $ &$ 0.0025 $ & $  0.0025$ \\
	$\{ 2,3,4\}$   & $ 0.0111 $ &$   0.0127 $ & $  0.0025  $ & $ 0.0025$ \\
	$\{ 4,5,6\}$   & $ 0.0278$ &$ 0.0278$&$0.0278 $ & $ 0.0278$ \\
	$\{ 6,7,8\}$   & $ 0.0025  $ &$   0.0025  $ &$   0.0127   $ &$  0.0111 $ \\
	$\{7,8,9 \}$   & $ 0 $ &$     0 $ &$     0.0004 $ &$      0.0004 $\\ 
	$\{6,8,9 \}$   & $ 0.0025   $ &$  0.0025  $ &$   0.0111    $ &$ 0.0127 $ \\
	\hline
\end{tabular}}
\end{center}
\end{table}

The sum of all the hyperedge cost or each column of Table \ref{tab:3uni_cosr} is 
$0.0569$. It can be noticed that the hyperedges among densely connected have less cost as compared to others. For example, hyperedge $\{1,2,3\}$ has less cost as compared to the hyperedge $\{4,5,6\}$. 

Such hyperedges with relatively smaller cost can be termed as ``smooth" hyperedge because they contain nodes which are densely connected by other hyperedges. For example, nodes 6 and 8 in Figure \ref{fig:ex2_3uni} are connected by 3 hyperedges. 
Ideally, the new hyperedge should be constructed among the nodes which are densely connected by other hyperedges. It is observed that the cost for such hyperedges is smaller compared to other hyperedges. So, we propose a hyperedge prediction algorithm which promotes the construction of hyperedges with minimal cost. The proposed algorithm is summarized in Table \ref{tab:algo}. 

\begin{table}[H]
	\caption{Hyperedge Prediction Algorithm }
	\label{tab:algo}
		\hrulefill
		{\normalsize
	\begin{enumerate}
		\item Construct the unnormalized or normalized tensor Laplacian as shown in \eqref{eq:Lt_ent} or \eqref{eq:Lt_ent_norm} respectively. 
		\item Compute the Fiedler eigenpair $(\lambda_{\star} ,\mathbf{v}_{\star})$ using \eqref{eq:fiedler_eqn}. 
		\item For a given set of potential hyperedges $E_p$, compute the construction cost using \eqref{eq:am_gm_main} and the Fiedler eigenvector computed in previous step. The same can be stated as $c_l = \{l_{e_j}(\mathbf{v}_{\star}) | e_j \in E_p \}  $.
		\item Prefer the construction of hyperedges with minimal construction cost. 
	\end{enumerate}}
\hrulefill
\end{table}

In this section, we proposed a novel hyperedge prediction algorithm using the spectral framework. In the next section, the working and efficacy of the proposed method are demonstrated using simple toy examples and real hypergraphs. 

\section{Experiments}
\label{sec:Exp}
We consider simple hypergraphs with interesting structural properties to validate the functioning of the proposed algorithm in section \ref{sec:ex1} and real hypergraphs later on. 
\subsection{Synthetic hypergraph: Example 1}
\label{sec:ex1}
We consider a 3-uniform symmetrical hypergraph with nine nodes and seven hyperedges shown in Figure \ref{fig:ex2_3uni}. The task is to predict the best new set of hyperedges to be formed from all the potential set of hyperedges.

We construct the tensor Laplacian and compute the eigenvalues and eigenvectors, as stated in the proposed algorithm described in  Table \ref{tab:algo}. We arrive at four Fiedler eigenvectors mentioned in \eqref{eq:ex1_eigvec} with same eigenvalue of $0.0569$.

All the above eigenvectors are then used for computing the cost of each potential hyperedges. To predict the best set of new hyperedges, all the potential hyperedges are considered. As there are $9$ nodes, one could have ${9 \choose 3} = 84$. Seven hyperedges are further removed as they already exist in the hypergraph, which leaves us with $84 - 7 = 77$ potential hyperedges.  The cost for each of these potential 77 hyperedges is computed using each of the $4$ eigenvectors mentioned in \eqref{eq:ex1_eigvec}. The next step is to rank these potential hyperedges based on the increasing order of their formation cost. 

It is observed that the cost computed from the first two eigenvectors (first two columns of $\mathbf{V}$ in \eqref{eq:ex1_eigvec}) are the same. The same holds for the other two eigenvectors. So, the preferential rank of new hyperedge formation from these two sets of eigenvectors is also same. Due to space constraints, only $10$ hyperedges with minimal formation cost are mentioned in Table \ref{tab:ex1_cost_unnorm}.

{\small
\begin{table}[H]
  \caption{Cost of new hyperedges using Unnormalized Laplacian}
  \label{tab:ex1_cost_unnorm}
  \begin{center}
  \begin{tabular}{c|c|l}
    \toprule
    $\text{hyperedges}_1$ & $\text{hyperedges}_2$ & cost \\
    \midrule
    $\{6,7,9\}$ & $\{1,3,4\}$& 0.0028\\
    $\{5,6,8\}$& $\{2,4,5\}$ & 0.0139\\
    $\{1,3,4\}$ & $\{6,7,9\}$& 0.0142\\
    $\{5,6,7\}$ & $\{1,4,5\}$ & 0.0152\\
    $\{5,6,9\}$ & $\{3,4,5\}$ & 0.0152\\
    $\{5,8,9\}$ & $\{1,2,5\}$ & 0.0195\\
    $\{5,7,8\}$ & $\{2,3,5\}$ & 0.0195\\
    $\{5,7,9\}$ & $\{1,3,5\}$ & 0.0205\\
    $\{3,4,5\}$ & $\{5,6,9\}$ & 0.0365\\    
    $\{1,4,5\}$ & $\{3,5,6\}$ & 0.0379\\    
  \bottomrule
\end{tabular}
\end{center}
\end{table}}

The cost of new hyperedges can also be calculated from the eigenvectors of normalized tensor Laplacian defined in \eqref{eq:Lt_ent_norm}. The same analysis can be performed 
using the normalized tensor Laplacian defined in to favor all nodes equally with respect to their degree distribution. The construction cost of new hyperedges using the eigenvectors of normalized tensor Laplacian is reported in Table \ref{tab:ex1_cost_norm}. 

{\small
\begin{table}[H]
  \caption{Cost of new hyperedges using Normalized Laplacian}
  \label{tab:ex1_cost_norm}
  \begin{center}
  \begin{tabular}{c|c|l}
    \toprule
    $\text{hyperedges}_1$ & $\text{hyperedges}_2$ & cost \\
    \midrule
    $\{6,7,9\}$ & $\{1,3,4\}$& $3.3 \times 10^{-4}$\\
    $\{2,3,5\}$& $\{5,8,9\}$ & 0.0142\\
    $\{1,2,5\}$ & $\{5,7,8\}$& 0.0160\\
    $\{1,3,5\}$ & $\{5,7,9\}$ & 0.0172\\
    $\{1,3,4\}$ & $\{6,7,9\}$ & 0.0173\\
    $\{3,4,5\}$ & $\{5,6,9\}$ & 0.0197\\
    $\{2,4,5\}$ & $\{5,6,8\}$ & 0.0254\\
    $\{4,5,7\}$ & $\{1,5,6\}$ & 0.0375\\
    $\{4,5,9\}$ & $\{3,5,6\}$ & 0.0375\\    
     $\{1,4,5\}$ & $\{5,6,7\}$ & 0.0386\\    
  \bottomrule
\end{tabular}
\end{center}
\end{table}}

Following observations can be made from Table \ref{tab:ex1_cost_unnorm} and Table \ref{tab:ex1_cost_norm}:
\begin{enumerate}
    \item The most obvious
    hyperedge to be formed for this hypergraph is $\{1,3,4\}$ and $\{6,7,9\}$. This can also be seen as nodes $1,2,3,4$ are densely connected with other hyperedges
    . So, the only remaining hyperedge among the four possible hyperedges is $\{1,3,4\}$. The same study holds for the hyperedge $\{6,7,9\}$. \\
    The most probable hyperedges are predicted by the proposed algorithm as it has minimum construction cost mentioned in first row of Table \ref{tab:ex1_cost_unnorm} and Table \ref{tab:ex1_cost_norm}. \\
    This trivial task of predicting the most probable hyperedge helps to validate the functioning of the proposed algorithm. 
    \item It can be observed that the most probable hyperedge is $\{1,3,4\}$ and $\{6,7,9\}$ using unnormalized and normalized Laplacian. However, the second best hyperedge is different. The probable hyperedge for unnormalized Laplacian is $\{2,4,5\}$ while it is $\{2,3,5\}$ for the normalized case. In both cases, nodes 2 and 5 are present.  Note that, node three is given more preference in the normalized case as compared to node 4. This behavior is expected 
    because the significance of nodes with a smaller degree will enhance after normalization compared to the unnormalized case. 
    
    Thus, this observation encourages the use of normalized Laplacian for hypergraphs having high variance in the degree distribution. \cite{von2007tutorial} also establishes a similar preference for using normalized or unnormalized Laplacian in case of graphs.  
\end{enumerate}

\subsubsection{Eigenvectors of Tensor vs. Matrix representation}
Most of the existing methods using matrix representation model the dyadic interaction among nodes only and further predict the hyperedges of cardinality greater than 2. To manifest the effectiveness of tensor eigenvectors, we propose a slight variation of the proposed algorithm.

One of the crucial steps of the proposed algorithm (in Table \ref{tab:algo}) is the computation of tensor eigenvectors which captures super-dyadic interactions. To demonstrate the importance of this step, we replace it with the computation of eigenvectors of graph Laplacian (matrix) derived from hypergraph reduction using \eqref{eq:inci_a}. All the other steps in the algorithm remain the same.

The construction cost of new hyperedges derived from the Fiedler eigenvector of reduced hypergraph is shown in Table \ref{tab:ex1_graph}. 

\begin{table}[H]
  \caption{Cost of new hyperedges using Normalized Laplacian of reduced hypergraph}
  \label{tab:ex1_graph}
  \begin{center}
  \begin{tabular}{c|l}
    \toprule
    $\text{hyperedges}_1$  & cost \\
    \midrule
    $\{1,5,7\}$ & 0 \\ 
    $\{1,5,9\}$ & 0 \\ 
    $\{2,5,8\}$ & 0 \\ 
    $\{3,5,7\}$& 0 \\ 
    $\{3,5,9\}$& 0 \\ 
    $\{2,5,9\}$& 0.0038 \\ 
    $\{2,5,7\}$ & 0.0038 \\
    $\{1,5,8\}$ & 0.0038 \\
    $\{3,5,8\}$ & 0.0038 \\
    $\{6,7,9\}$ & 0.0217 \\
  \bottomrule
\end{tabular}
\end{center}
\end{table}
It can be easily stated that the above results are not as expected and do not capture the interaction among three nodes. This can be justified theoretically as the Laplacian cost function is a second order homogeneous polynomial modeling dyadic interaction only whereas the tensor-based Laplacian cost function is a third order homogeneous polynomial capturing the super-dyadic interactions. 

In this example, we investigated various features of the proposed algorithm such as 
\begin{enumerate}
    \item Deriving preferential order of new hyperedges.
    \item Behaviour of normalized and unnormalized Laplacian.
    \item Effectiveness of the tensor eigenvectors in capturing the super-dyadic interactions.
\end{enumerate}

\subsection{Real Hypergraphs}
In this subsection, we analyze the performance of the proposed algorithm on real hypergraphs. We first describe the datasets, baselines, and then the experimental settings used to evaluate these hyperedge prediction baselines. 

\subsubsection{Datasets}
We consider five datasets with varying number of nodes and hyperedges from different domain. We have mentioned the size of largest of connected component consisting hyperedges of cardinality 3 in Table \ref{tab:real_data}. Please note that we have performed the experiments and shown results for 3-uniform hypergraphs for simplicity but the proposed approach can be applied to any $k$
hypergraph. 

{\small
\begin{table}[H]
  \caption{Datasets}
  \label{tab:real_data}
  \begin{center}
  \begin{tabular}{l|c|c|c}
    \toprule
    Name & $|V|$ & $|E_3|$ & Reference \\
    \midrule
    uchoice Bakery & 50 & 24674 &   \cite{benson2018discrete} \\ 
    uchoice Walmart Dept  & 66 & 24365 &  \cite{benson2018discrete} \\ 
    contact-primary school & 242 & 9262 &   \cite{stehle2011high}\\ 
    contact-high school & 317 & 7475 &   \cite{mastrandrea2015contact}\\ 
    NDC-Substances & 570 & 6327 &  \cite{benson2018simplicial}\\ 
  \bottomrule
\end{tabular}
\end{center}
\end{table}}

These datasets were constructed in following manner: 
\begin{enumerate}
    \item uchoice Bakery: Nodes represent the items in bakery and hyperedges are constructed among the items bought together. 
    \item uchoice Walmart Dept: Nodes represent the ``department'' of an item in the shop and a hyperedge is constructed among the department whose items were co-bought. 
    \item contact-primary school \& contact-high school: Nodes are people in the corresponding school and hyperedges are constructed among the people if they interacted with each other in interval of 20 seconds. The interaction was recorded by a wearable sensor. 
    \item NDC-Substances: The data is taken from US National Drug Code (NDC), where a hyperedge denotes a drug and the nodes represent the substances used in that drug. 
\end{enumerate}
We further briefly discuss the existing hyperedge prediction approaches. 

\subsubsection{Baselines}
We consider some of the most widely used hyperedge prediction baselines in this subsection. Every method tries to construct a ``similarity score'' of the potential hyperedge by its model. A large similarity score of a potential hyperedge indicates that it is more likely to be formed as compared to potential hyperedges with low similarity score. This similarity score is used to as a proxy to predict the new hyperedges. Hence, we discuss the approach in which each of the following baselines construct that similarity score: 
\begin{enumerate}
    \item Common Neighbours (CN) \cite{newman2001clustering}:  For a potential hyperedge, the similarity score is the sum of number of common neighbours of two nodes taken at a time in the given hypergraph. It should be noted that similarity score is computed using the local information in this approach. 
    \item Katz \cite{katz1953new}: The similarity score is computed based on the global information using the paths connecting the two nodes. For a hyperedge with $m$ nodes, we consider all the possible $k(k-1)/2$ pairs of nodes. 
    \item HPRA \cite{kumar2020hpra}: This is a recently proposed algorithm which computes the similarity score by extending the use of resource allocation approach \cite{zhou2009predicting, lu2011link} from graphs to hypergraphs. This method also proposes a modified hypergraph reduction method which preserves the node degrees of original hypergraph in the resulting graph \cite{kumar2020hypergraph}. 
\end{enumerate}
We further describe the experimental settings used in the evaluation of these methods. 

\subsubsection{Experimental settings}
The first step before applying any of the above methods is to construct a potential set of hyperedges. The naive approach of considering all possible hyperedges can not be used due to the large number of potential hyperedges unlike the case of small synthetic hypergraphs. We used that approach in Section \ref{sec:ex1} to show the functioning of the proposed method. 

The first step is to remove a few existing hyperedges from the given hypergraph. A hyperedge prediction algorithm is then evaluated on the basis of predicting the removed hyperedges. A good algorithm should also not predict the non-existing hyperedges in original hypergraph. So, we construct our test set (or potential set) of hyperedges by considering both the removed and non-existing hyperedges. The number of removed hyperedges in our experiments is maintained as $10\%$ of the existing number of hyperedges. The rest $90 \%$ of the hyperedges are used for training. 

As the non-existing hyperedges are considered in test set, the choice of non-existing hyperedges plays a vital role in the evaluation of hyperedge prediction algorithms. We utilize recently proposed negative sampling approach \cite{patil2020negative} to construct the set non-existing hyperedges. The first step of this negative sampling approach is to reduce the hypergraph to a graph and then connect the neighbors of nodes in a randomly sampled edge to the chosen edge in order to construct the hyperedge. This process is repeated until the desired number of non-existing hyperedges are sampled, which we choose to be $3$ times the number of existing hyperedges in the given hypergraph for training. Please note that this approach can finally provide a hyperedge that is already existing in the hypergraph, whereas our motive was to sample non-existing hyperedges. So in this work, we remove those hyperedges from this ``non-existing set'' of hyperedges which already existed in the original hypergraph in order to have a proper evaluation of hyperedge prediction algorithms.  

\subsubsection{Results}
We run the experiments on each dataset at least 20 times, removing $10\%$ of hyperedges randomly in each run. The quality of predicted hyperedges by any algorithm is compared by using average F1 score \cite{yang2013overlapping}. A higher average score indicates that the performance of the corresponding algorithm is better. The mean of average F1 scores computed from the $20$ runs on each dataset are presented in Table  \ref{tab:results_data}. 

{\small
\begin{table}[H]
  \caption{Results}
  \label{tab:results_data}
  \begin{center}
  \begin{tabular}{l|c|c|c|c||c}
    \toprule
    Dataset & CN & Katz & HPRA & Proposed & PI \\
    \midrule
    uchoice Bakery & 0.3445 & 0.3457 & 0.3459 & 0.3658 & 5.53 \\ 
    uchoice Walmart Dept & 0.2592 & 0.2671 & 0.2623 & 0.3322 & 23.4 \\ 
    contact-primary school & 0.2150 & 0.2164 & 0.2242 & 0.2738 & 21.51 \\ 
    contact-high school & 0.2140 & 0.2163 & 0.2226 & 0.2794 & 24.19 \\ 
    NDC-Substances & 0.1743 & 0.1779 & 0.1808 & 0.2343 & 24.42 \\ 
  \bottomrule
\end{tabular}
\end{center}
\end{table}}
													
In order to compare the performance of all algorithms, we also define the relative performance improvement (PI) as: 
\begin{align}
    \text{PI} = \frac{\text{Avg-F1}_{\text{prop}} - \text{Avg-F1}_{\text{base}} } {\text{Avg-F1}_{\text{base}}} \times 100
\end{align}
where $\text{Avg-F1}_{\text{prop}}$ denotes the average F1 score by proposed algorithm and $\text{Avg-F1}_{\text{base}}$ denotes the average F1 scores of the best baseline algorithm for the corresponding dataset. We compute PI score for each of the 20 runs separately and present the mean of those $20$ runs in last column of Table \ref{tab:results_data}. A positive PI score indicates that the proposed algorithm has performed better and its magnitude signifies the improvement. 

It is clearly evident that the proposed algorithm has outperformed the existing baselines by a considerable margin from Table \ref{tab:results_data}. In this section, we discussed the performance of proposed method on real datasets. We make concluding remarks and provide directions for future work in the next section. 

\section{Conclusion \& Future Work}
\label{sec:concl}
In this article, we proposed a novel framework for hyperedge prediction for k-uniform hypergraphs. The critical challenge for this task was modeling complex interactions among multiple nodes. We utilized the tensor-based representation of hypergraphs, which helps to model the super-dyadic interactions among the nodes. The proposed algorithm prefers to construct the hyperedges with minimal construction cost. In the perspective of spectral hypergraph theory, this can also be perceived as the inclusion of new hyperedges such that there is minimal perturbation in the ``smoothness'' of the hypergraph. The functioning and fruitful merits of the proposed algorithm were demonstrated using synthetic and real hypergraphs. The future directions of this work are along the lines of performing a similar analysis for non-uniform and directed hypergraphs. 


\section*{Acknowledgements} 
This work was partially supported by Intel research grant RB/18-19/CSE/002/INTI/BRAV to BR.

\bibliographystyle{spmpsci}      

\bibliography{sample-base}

\end{document}